\documentclass[useAMS,usenatbib]{mn2e}
\usepackage{graphicx}
\usepackage{graphics}
\usepackage{epstopdf}
\usepackage{float}
\usepackage{bm}
\usepackage{epsf}
\usepackage{url}
\usepackage{color}
\usepackage{natbib}
\usepackage{amsmath}

\makeatletter
\renewcommand\@afterheading{%
  \@nobreaktrue
  \everypar{%
    \if@nobreak
      \@nobreakfalse
      \clubpenalty 1
      \if@afterindent \else
        {\setbox\z@\lastbox}%
      \fi
    \else
      \clubpenalty 1
      \everypar{}%
    \fi}}
\makeatother

\newcommand{\kms}{\,km s$^{-1}$}

\definecolor{ForestGreen}{rgb}{0.3,0.7,0.3}

\title[The Unusual Milky Way-Local Sheet System]{The Unusual Milky Way-Local Sheet System: \\ Implications for  Spin Strength and Alignment}
\author[M.A. Aragon-Calvo]{M.A. Aragon-Calvo$^{1}$, Joseph Silk$^2$,  Mark Neyrinck$^{3,4,5}$\\
$^1$Instituto de Astronomía, UNAM, Apdo. Postal 106, Ensenada 22800, B.C., México\\
$^2$Institut d’Astrophysique de Paris, UMR7095:CNRS UPMC-Sorbonne University, F-75014, Paris, France\\
$^3$ Ikerbasque, Basque Foundation for Science, 48011 Bilbao, Spain\\
$^4$ Department of Physics, University of the Basque Country UPV/EHU, 48080 Bilbao, Spain\\
$^5$ DIPC, Basque Country UPV/EHU, 20018 San Sebasti\'{a}n, Spain
}

\begin{document}

\date{}

\pagerange{\pageref{firstpage}--\pageref{lastpage}} \pubyear{2002}
\maketitle
\label{firstpage}

\begin{abstract}

The Milky Way and the Local Sheet form a peculiar galaxy system in terms of the unusually low velocity dispersion in our neighbourhood and the seemingly high mass of the Milky Way for such an environment. 
Using the TNG300 simulation we searched for Milky Way analogues (MWA)  located in cosmological walls with velocity dispersion in their local Hubble flow similar to the one observed around our galaxy.
We find that MWAs in Local-Sheet analogues are rare, with one per (160-200 Mpc)$^3$ volume.
We find that a Sheet-like cold environment preserves, amplifies, or simplifies environmental effects on the angular momentum of galaxies. In such sheets, there are particularly strong alignments between the sheet and galaxy spins; also, these galaxies have low spin parameters. These both may relate to a lack of mergers since wall formation. We hope our results will bring awareness of the atypical nature of the Milky Way-Local Sheet system. Wrongly extrapolating local observations without a full consideration of the effect of our cosmic environment can lead to a \textit{Copernican bias} in understanding the formation and evolution of the Milky Way and the nearby Universe.

\end{abstract}
\begin{keywords}
Cosmology: large-scale structure of Universe; galaxies: kinematics and dynamics, Local Group; methods: data analysis, N-body simulations
\end{keywords}

\section{Introduction}

The Milky Way (MW) is an SBb/c  galaxy with an estimated mass between $8.3\times10^{11}$M$_\odot$ and $2 \times 10^{12}$ M$_\odot$ \citep{Karukes20}. It is located near the center of a flat sheet of galaxies extending up to 4 Mpc along along the supergalactic plane \citep{Kroupa05,Metz07,Pawlowski13} with a density close to the mean density of the Universe \citep{Klypin15} and delineated by a ring of luminous galaxies, the ``Council of Giants'' \citep{McCall14}. The ``Local Sheet'' in which  our galaxy is located belongs to a class of cosmic structures known as cosmic walls. It is bounded to the supergalactic north and south by the \textit{Local Void} and the \textit{Southern Void} respectively \citep{Tully08}. 
There are no galaxies in the line of sight between the our galaxy and the center of both north and southern voids. The Milky Way is sitting at the edge of a cosmic cliff.

An increasing body of evidence points to the Milky Way and its immediate cosmic environment as somewhat uncommon occurrences in the Universe. The velocity dispersion measured for galaxies located within 3 Mpc from our own galaxy is $\sigma_v \sim 25-40$ \kms \citep{Karachentsev02,Karachentsev03}, one order of magnitude lower than predictions \citep{Peebles80,Governato97} and observed pairwise velocity dispersions \citep{Landy02,Zehavi02,Scoccimarro04}. The ``coldness'' of the local Hubble flow seems to be a global property of the Local Sheet as indicated by observations around the nearby Centaurus A / M83 system, showing a dispersion of $\sigma_v \sim 32$ km s$^{-1}$ \citep{KarachentsevCenA07}.
The coldness of the local Hubble flow can be explained as the result of the characteristic geometry and dynamics of cosmic walls \citep{Aragon11}. The central regions of walls are less dense than their surrounding denser ``ring" of filaments, producing a local expansion on top of the global Hubble flow which damps peculiar velocities.

The halo mass function in walls also indicates the peculiarity of Milky Way Analogue (MWA) galaxies as it is roughly one order of magnitude lower than the global mass function in the MW mass range (\citet{AragonThesis07,Punyakoti18}, see also \citet{Neuzil20}). The rarity of MWA galaxies is evident, albeit not explicitly addressed, in studies of MWA in which constraints on halo mass, local density and geometry are included and also in constrained realizations of the Local Group. In both cases a large volume or a large number of realizations are needed in order to find MWA candidates (see for instance \citet{Garrison14} and \citet{Fattahi16}).

Walls are dynamically young structures, being the result of the gravitational collapse along one axis \citep{Zeldovich70}.
In the tidal torque theory \citep{Hoyle49,Peebles69,White84}, the tidal field governs spin. In a collapsing wall environment, this tidal field squashes protohaloes into a plane. Any aspherical lumps on the protohalo will get caught in the tidal field, imparting torque to the halo, with its spin tending to end up perpendicular to the axis of collapse \citep[see also][]{NeyrinckEtal2020}. 
This alignment between galaxy spin axes and their parent wall planes is small but statistically significant in dark matter computer simulations \citep{Aragon07,Hahn07}. Locally, the alignment of edge-on spiral galaxies (including our own) perpendicular to the Local Sheet is so strong that it can be seen by eye. As shown in \citet{Navarro04}, gas/stars can retain their primordial orientation better than dark matter, due to the latter being more affected by mergers. The strength of the alignment signal in our neighborhood may be the result of disk galaxies keeping their original spin alignment and a quiet mass growth and low velocity dispersion in the relatively low density of the Local Sheet. On the other hand, mergers are a well known mechanism for angular momentum acquisition \citep{Vitvitska02}. In the case of galaxy interactions within walls, mergers would tend to induce a spin aligned with the normal of the wall, not with the plane, as prescribed by the TTT.


\begin{figure}
  \centering
  \includegraphics[width=0.48\textwidth,angle=0.0]{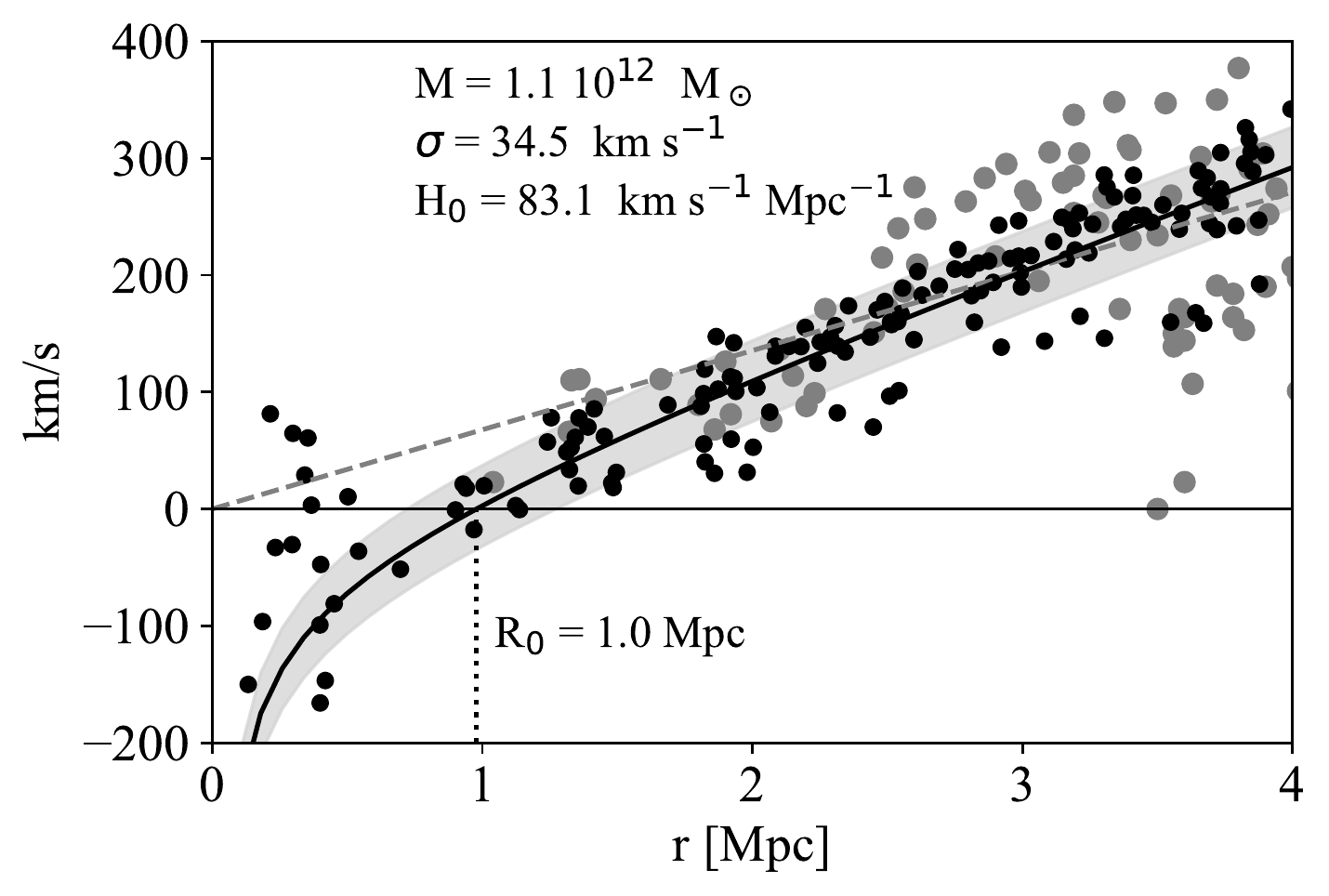}
  \caption{Hubble diagram measured from a MWA in our sample. The dots correspond to background galaxies inside the search radius, the solid line indicates the best fit to the model given in sec. \ref{sec:measuring_hubble}. The peculiar velocity dispersion is indicated by the dashed lines. The vertical dotted line marks the radius of zero velocity. We also show the unperturbed Hubble expansion $v=H r$ as the dashed gray line. The mass and fitting parameters are shown in the figure. Gray dots correspond to real galaxies presented in \citet{Karachentsev03}.}
  \label{fig:hubble-example}
\end{figure}

\section{N-body Simulation and LSS analysis}\label{sec:N-body_halos}

The analysis presented here is based on Illustris TNG300 \citep{Nelson18}, a cosmological hydrodynamical simulation of galaxy formation enclosed in a box of 205$h^{-1}$Mpc of side containing dark matter and gas particles with masses of $5.9\times10^7$ and $1.1\times10^7$ $h^{-1}$M$_{\odot}$ respectively. The cosmological parameters used are given by Planck results \citep{Planck16}. While TNG300 has a relatively low mass resolution it has the large volume needed to sample rare cold Local Sheet environments as we will see in the following sections.

We defined two set of haloes: background haloes used to sample the velocity field and Milky Way Analogues. The background halo sample consisted of subfind haloes in the mass range  $1\times 10^{9} < $M$ < 5\times 10^{12}$ M$_{\odot}$ and $V_{\textrm{\tiny max}} > 20$ km s$^{-1}$. MWAs were defined as central haloes
with $V_{\textrm{\tiny max}} > 20$ km s$^{-1}$ in the mass range $10^{12} < $ M $< 2\times10^{12}$ M$_\odot$. We extracted mass accretion histories for the MWA sample and identified merger events as increases in $30 \%$ of the total mass of the halo between snapthots.
The MWA sample was classified into spiral and ellipticals using a custom convolutional neural network feed with synthetic images created from the stellar component of the halos observed in the $r,g,i$ SDSS filters (Aragon-Calvo, in preparation, see for instance \citet{Rodriguez19}).
For every MWA halo we computed the spin parameter following the definition of \citet{Bullock01}:
$\lambda = \vert J \vert / ( \sqrt{2} M_{200} R_{200} V_{200} )$,
where $J$ is the angular momentum, $M_{200}$ and $R_{200}$ are the virial mass and radius respectively and $V_{200}$ the circular velocity.

%

The density field used for cosmic web analysis was computed by running a resimulation of TNG300 sampled on a regular grid with $512^3$ particles. The initial conditions were sharp-$k$-smoothed in order to target structures larger than 4 h$^{-1}$Mpc as described in \citet{Aragon10b} and evolved to $z=0$ using the GADGET-2 N-body code \citep{Springel05}. From the final snapshot we labeled wall regions using the Hessian-based MMF method \citep{Aragon07b,Aragon14}.

%
\subsection{Measuring the Hubble flow}\label{sec:measuring_hubble}

The Hubble flow around MWAs was measured by first selecting background haloes inside a shell of $1<r<4$ Mpc centered in the target MWA galaxy. This distance range was chosen to roughly follow the range used in real measurements in the local Universe \citep{Karachentsev09} although extending to a larger radius in order to increase the number of available galaxies. Only MWAs with more than 10 neighbours were included in the analysis. 
We computed velocity dispersions $\sigma_v $ from the model-subtracted radial velocities centered on the MWA using the model \citep{Peirani08}: $v(r) = H_0 r - H_0 R_0 ( R_0/r)^{1/2}$, 
where $r$ is the distance from the MWA and $H_0$ and $R_0$ are the fitted Hubble constant and radius of the  zero-velocity surface respectively (see Fig. \ref{fig:hubble-example}).

\begin{figure}
  \centering
  \includegraphics[width=0.47\textwidth,angle=0.0]{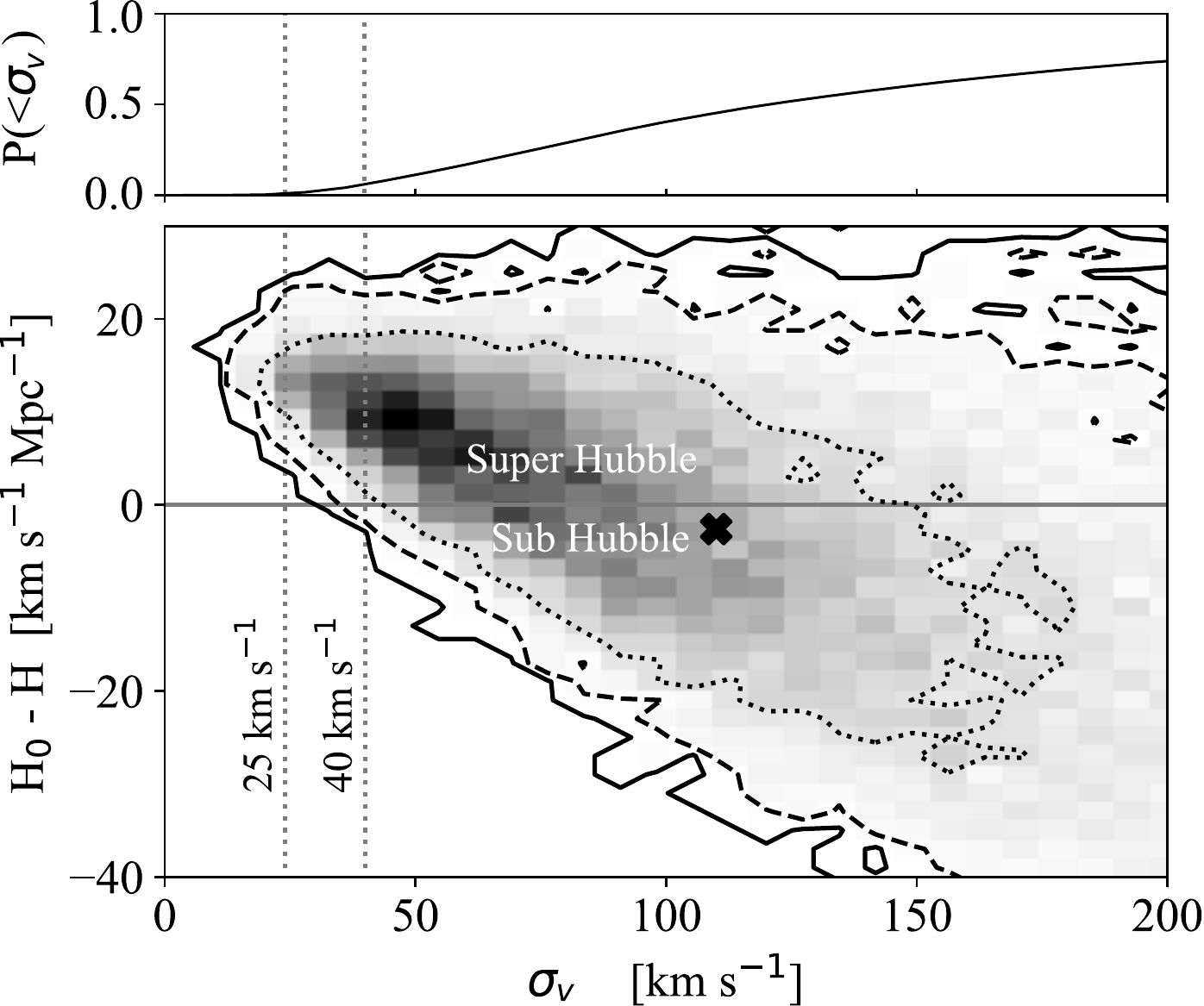}
  \caption{Hubble flow excess ($H_0-H$) as a function of $\sigma_v$ around the MWA sample. The vertical dotted lines indicate the range of observed velocity dispersions around our galaxy (see text for details). The 2D distribution is shown as a gray background with inverted color scale. The black cross shows the median of $\sigma_v$ and $H-H_0$ The contours enclose the areas containing 86,95 and 99.7 percent of galaxies. The top subplot shows the cumulative distribution of $\sigma_v$.}
  \label{fig:hubble_local_disp}
\end{figure} 

\section{Results}

Figure \ref{fig:hubble_local_disp} shows the distribution of $\sigma_v$ and Hubble flow excess (H$_0$-H) measured from the MWA sample. The distribution of both $\sigma_v$ and H$_0$-H have a roughly log-normal character. The peak of the velocity dispersion is surprisingly low, around $\sigma_v \sim 50$km s$^{-1}$. However our MWA sample is restricted to relatively massive galaxies and does not include galaxies in pairs, groups or clusters which can have a much higher $\sigma_v$. The tail of the distribution, being $\sim$log-normal, extends to much higher values in $\sigma_v$ and the mean of the distribution is $178$ \kms. The peak in the Hubble excess distribution is located at H$_0$-H $\sim$ 10 \kms. The tail of the distribution extends into the negative hundreds (not shown in Fig.  \ref{fig:hubble_local_disp} for clarity). The median value of H$_0$-H is -2.2 \kms, close to the global expansion.
The reported range in the velocity dispersion $25 < \sigma_v < 40$ corresponds to  $3\sigma - 1\sigma$ events for H$_0$-H=0. Interestingly, from Fig. \ref{fig:hubble_local_disp} we should expect to measure a Hubble flow excess around our galaxy of $\sim H+(10-15)$ km s$^{-1}$ Mpc$^{-1}$. 

\begin{figure}
  \centering
  \includegraphics[width=0.49\textwidth,angle=0.0]{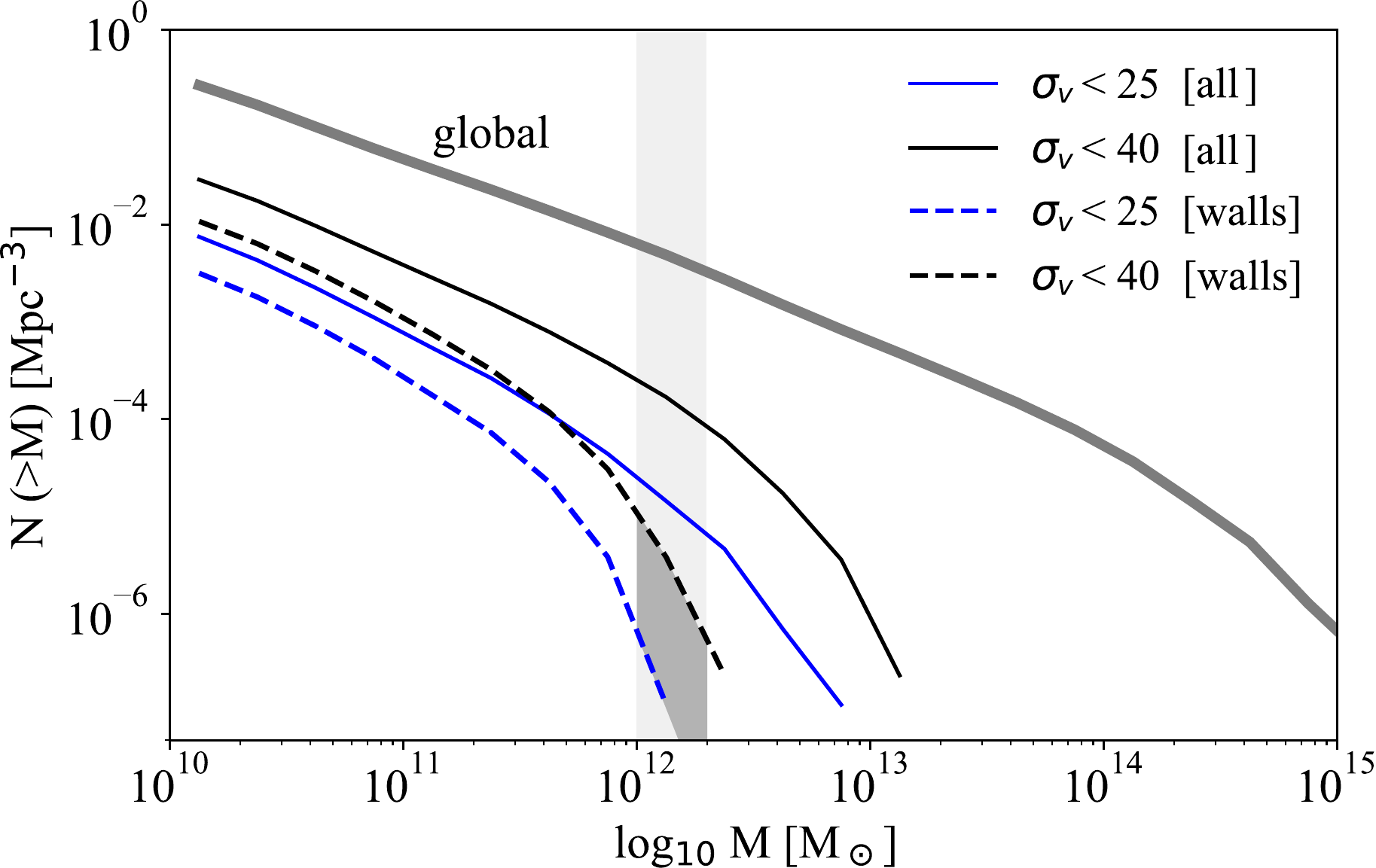}
  \caption{Mass function in different cosmic environments defined by their $\sigma_v$ and LSS morphology.  We show the mass function of haloes in cold ($\sigma_v < 25$ km s$^{-1}$) and warm ($\sigma_v < 40$ km s$^{-1}$) environments. For each $\sigma_v$ we also show the mass function of haloes in cosmic walls. The vertical light shaded area shows the mass range of MWA and the dark grey area highlights the most likely region where the MW would be located in the diagram. For comparison we show the global mass function.}
  \label{fig:mass_function_total} 
\end{figure} 

\begin{figure*}
  \centering
  \includegraphics[width=0.99\textwidth,angle=0.0]{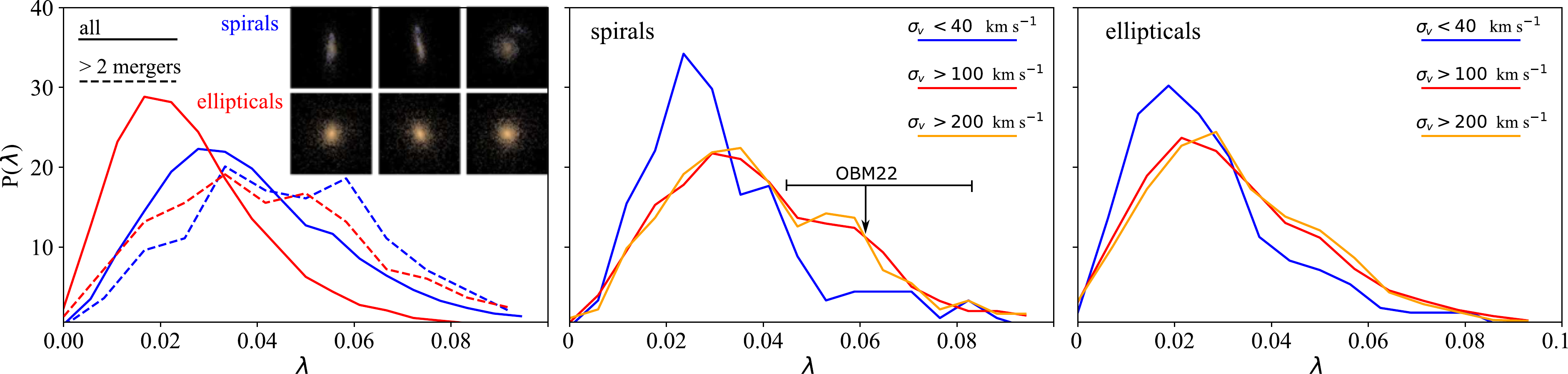}
  \caption{Distribution of halo spin parameter $\lambda$ for different MWA sub-samples. {\bf Left panel}: Distribution of $\lambda$ for spirals and ellipticals, the solid line corresponds to the full MWA sample and the dashed line to MWA galaxies with 3 or more major merger events since $z=2$. We also show synthetic images (three orthogonal orientations) of one spiral and one elliptical galaxy. 
{\bf Center panel}: The distribution of $\lambda$ for spiral galaxies in cold ($\sigma_v < 40$ km s$^{-1}$) and increasingly hot environments ($\sigma_v > 100$ km s$^{-1}$ and $\sigma_v > 2
00$ km s$^{-1}$). The vertical arrow and error bar indicate the most recent estimate of the spin parameter for the Milky Way ($\lambda=0.061^{+0.022}_{-0.016}$, \citet{Obreja22} (OBM22)).
{\bf Right panel}: Same as the center panel but for elliptical galaxies.}
  \label{fig:hubble_spin_dist}
\end{figure*}

%
\subsection{How common are MWAs in cold walls?}

Figure \ref{fig:mass_function_total} shows the effect of cosmic environment, defined in terms of geometry (walls) and $\sigma_v$, on the halo mass function (see also \citet{AragonThesis07,Punyakoti18}). The mass function decreases in height and slope with decreasing Hubble velocity dispersion. The effect of environment is even stronger when we consider only haloes in walls. At low masses the mass function in all environments is a scaled version of the global mass function. The slope in the mass function begins to decrease with increasing mass around $10^{12}$ M$_{\odot}$. The ``knee" of the curve associated with a characteristic mass in cold environments is more than one order of magnitude lower than the ``knee" of the global population. The slope in the mass function in cold wall environments is significantly steeper than the global population, the effect being even more pronounced after the characteristic mass range. The Milky Way is right at the border of the knee in the mass function in walls in the two regimes of $\sigma_v$ considered here. 
The probability of drawing a MWA in cold wall environments from the general population $P(>M) = (N(>M)_{\textit{\tiny env}} / N(>M)_{\textit{\tiny global}}$ lies somewhere between 0.2\%  and 0.001\% for the ranges in mass and $\sigma_v$ considered here. This is equivalent to finding one MWA in a box of side $\sim 160$ and $\sim 200$ Mpc respectively. If we take the lowest estimate of $\sigma_v \sim 25$ \kms Milky Way analogues in cold Local Sheet environments can be considered extremely rare events. 


%
\subsection{Galaxy/Halo Spin in Local Sheet}

\subsubsection{Spin parameter}

Before addressing environmental effects, we study the spin parameter distribution of galaxies separated only by their morphology.
Figure \ref{fig:hubble_spin_dist} (a) shows that spirals have in general a higher $\lambda$ compared to ellipticals with the peaks in their $\lambda$ distribution, $P(z)$, at $\lambda \sim 0.03$ and $\lambda \sim 0.015$ respectively (see also \citet{Du21,Rodriguez22}). Note that we use a strict definition of morphology, rejecting galaxies with $<90\%$ probability of being a spiral/elliptical given by the neural network. Both spirals and ellipticals with 3 or more major-merger events since z = 1 have a higher spin parameter, supporting the idea of angular momentum acquisition via mergers \citep{Vitvitska02}.

The effect of environment on $\lambda$ is shown in Fig. \ref{fig:hubble_spin_dist} (b). Both spirals and ellipticals in walls have lower $\lambda$ in cold environments. The peak in $P(\lambda)$ increases with $\sigma_v$ and the trend  saturates for values higher than $\sigma_v \sim 100$ \kms. In all cases spirals have a higher $\lambda$ compared to ellipticals in their respective environment. 
Importantly, $P(\lambda)$ for galaxies in cold and hot environments not only differ in the position of their peaks but also in their character. The $P(\lambda)$ of galaxies in cold walls has a more narrow peak centered at a lower $\lambda$ value and has a lower high-$\lambda$ tail indicating a systematic lack of high-$\lambda$ galaxies in this environment. This effect is more pronounced in the case of spirals.


\subsubsection{Spin alignment}

Finally, we show the alignment of galaxy spins with walls they inhabit. Figure \ref{fig:spin-alignmnet} (right panel) shows that the alignment signal increases with decreasing $\sigma_v$ and increasing halo mass. 
The alignment signal of haloes with mass $<2\times10^{12}$ is weak, consistent with previous findings \citep{Aragon07,Hahn07} and showing a weak dependence with environment. In comparison, haloes above this mass present a significant alignment signal and dependence with environment. The star/disk component of galaxies is more strongly aligned than the dark matter component. Regarding cosmic environment, galaxies in cold environments are more strongly aligned than galaxies in hot environments. The strength in the alignment of the stellar component of galaxies in cold walls, similar to our Local Sheet, is very close to the observed value reported in \citet{Navarro04}. Interestingly, the Milky Way has a somewhat high $\lambda$ for its environment (see arrow in Fig.\ \ref{fig:spin-alignmnet} center panel). We note that the high $\lambda$ galaxies are also the strongest aligned. However the sample size is too small to draw a firm conclusion and report here. We will study this in a future work.

\begin{figure}
  \centering
  \includegraphics[width=0.4\textwidth,angle=0.0]{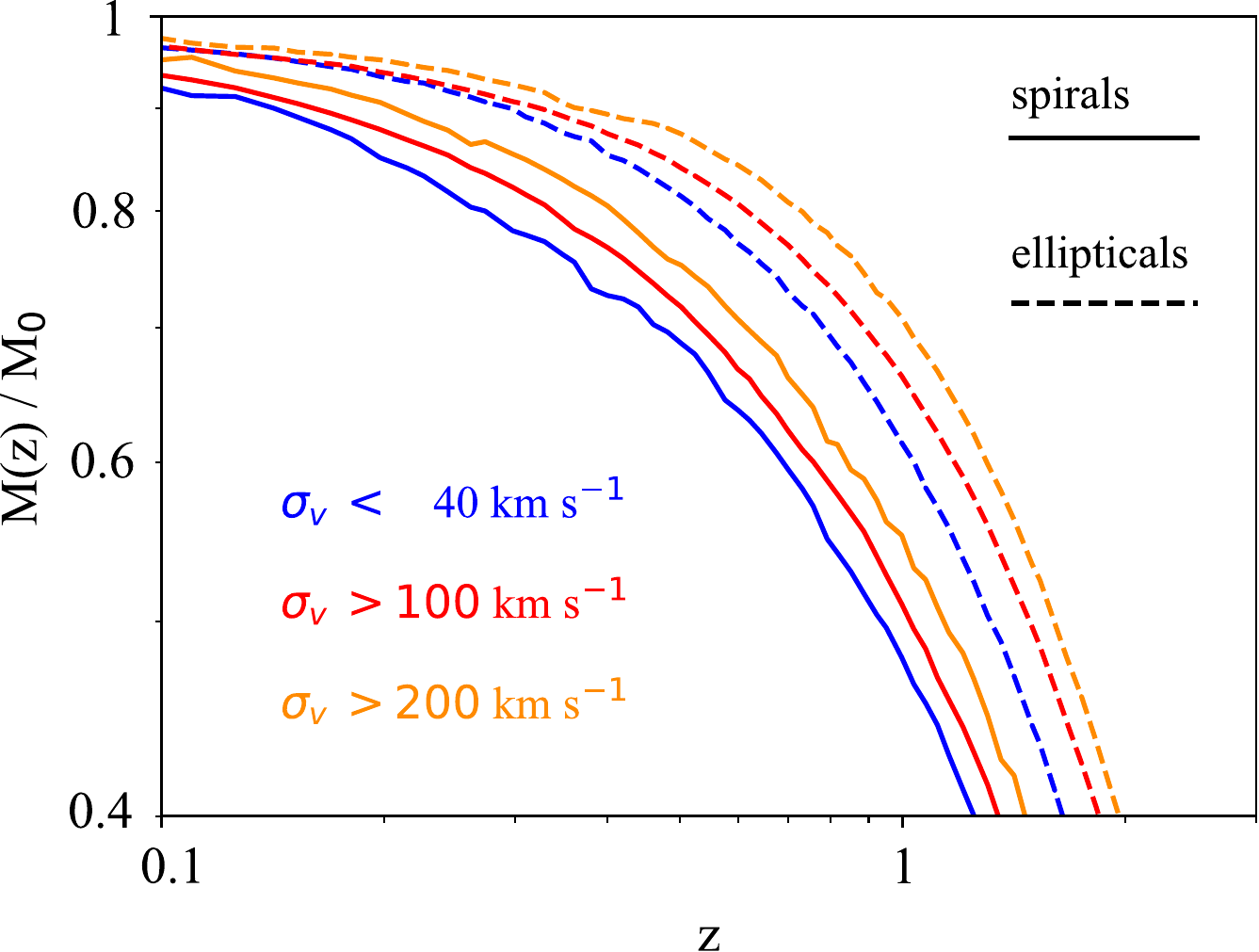}
  \caption{Mean mass accretion histories for spiral and elliptical MWA galaxies in increasingly hotter environments.
}
  \label{fig:spin-alignmnet}
\end{figure} 

\section{Conclusions and discussion}

In this work we present evidence that the Milky Way, a common galaxy by mass with respect to the global population, is an outlier when we consider the environment in which it is located. Improving on previous works, we describe the Local Sheet environment in terms not only of its geometry, but also its dynamics. We find that a Local Sheet and Milky Way environment, consisting of an unusually massive galaxy in a cold wall, is rare, occurring only once in a 160-200 Mpc box. We note that given the uncertainties on the mass of our galaxy, the measured $\sigma_v$ and the general size and shape of the Local Sheet, our results should only be taken as an approximate indication of the rarity of such systems. Adding conditions to a situation will always eventually give a vanishingly small probability so we adopt somewhat conservative constraints and define the MW-Local Sheet system in terms of only three parameters: LSS geometry, LSS dynamics and halo mass. Our results apply only to Milky Way analogues, if we consider the more massive Milky Way - Andromeda system, with a combined mass of $\sim 3-4\times 10^{12}$ M$_{\odot}$, such event is even rarer, so we do not analyze the joint situation of two large galaxies in a cold wall. But note that after the two galaxies merge, the situation will be the same as we analyze here, except with a larger combined mass.
A cold wall with a galaxy of this larger mass would be extremely rare; a much larger box than the TNG300 volume we analyze would be necessary to obtain an example.

We show that both spiral and elliptical galaxies located in cosmological walls with low $\sigma_v$ have a lower spin parameter, characterized by a narrower peak in the distribution located at lower $\lambda$ and a lack of high-$\lambda$ galaxies. This could indicate a lack of mergers in the quiet wall environments \citep{Hellwing21} (see also Fig. \ref{fig:hubble_spin_dist}) emphasized by the low $\sigma_v$.
At the same time, we found a strong alignment signal between the spin of spiral galaxies (selected using a neural network) and the plane of their host wall. The value of the alignment strength is close to the observed value around our galaxy. To our knowledge this is the strongest spin alignment in walls reported in $N$-body simulations. This is significant since we observe an unusually strong spin alignment of spiral galaxies around our galaxy, yet no previous study to date has been able to reproduce it. 

Our results highlight the importance of carefully characterizing the environment around our galaxy. The effect of the geometry and coldness of Local Sheet environment on angular momentum processes may help us better understand current problems in galaxy formation such as the co-rotating plane of satellites around M31 \citep{Ibata13}, the missing satellite problem \citep{Klypin99} among others.

\begin{figure}
  \centering
  \includegraphics[width=0.45\textwidth,angle=0.0]{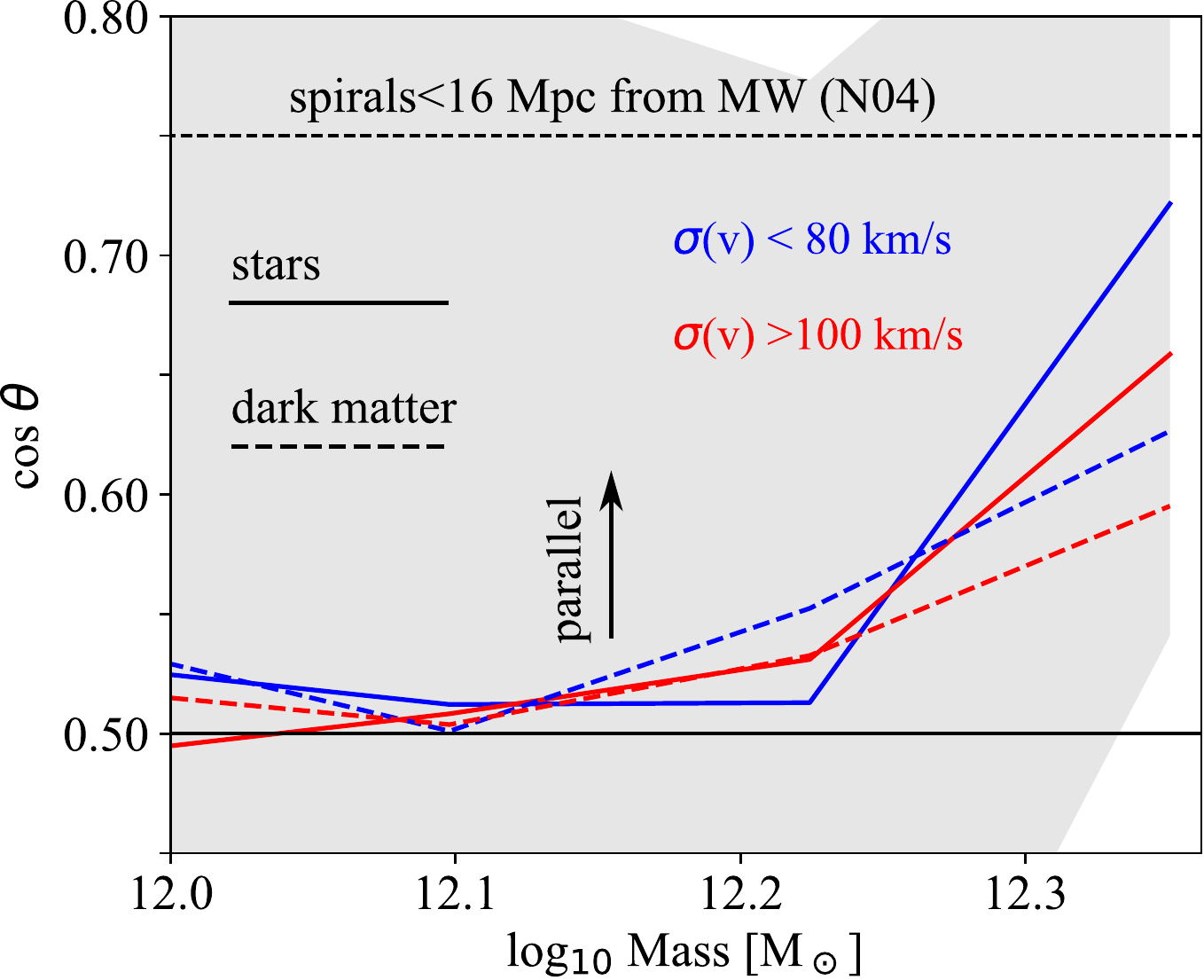}
  \caption{Spin alignment of spiral galaxies in walls. We include galaxies with 95\% probability of being spiral. The alignment $\cos\theta$ corresponds to the angle between the spin vector of the galaxy/halo and the plane of its host wall. We compute the alignment for the stars and dark matter components and cold and hot environments. The gray area indicates the dispersion inside each bin for the spin alignment of the star component of galaxies in cold environments. The other cases have similar dispersion. 
}
  \label{fig:spin-alignmnet}
\end{figure}

\section{Data Availability}

The data underlying this article were accessed from the Illustris TNG datababse (https://www.tng-project.org). The derived data generated in this research will be shared on reasonable request to the corresponding author.

\vspace{-3mm}
\section{Acknowledgements}
This research was partially funded by the UNAM-PAPIIT grant IA102020. M.A Aragon-Calvo would like to thank Dylan Nelson for kindly providing the TNG300 parameter file used for generating ICs. 

\vspace{-3mm}

\bibliography{refs} 
\bibliographystyle{mn2e}   

\end{document}